\documentclass[preprint,prd,nofootinbib,tightenlines,amsmath]{revtex4}
\usepackage{enumerate}
\usepackage{multirow}

\usepackage{epsfig}
\usepackage{graphics,color}      
\usepackage{verbatim}
\usepackage{amsmath}    
\catcode`\@=11
\def\sla@#1#2#3#4#5{{%
 \setbox\z@\hbox{$\m@th#4#5$}%
 \setbox\tw@\hbox{$\m@th#4#1$}%
 \dimen4\wd\ifdim\wd\z@<\wd\tw@\tw@\else\z@\fi
 \dimen@\ht\tw@
 \advance\dimen@-\dp\tw@ \advance\dimen@-\ht\z@
 \advance\dimen@\dp\z@
 \divide\dimen@\tw@ \advance\dimen@-#3\ht\tw@
 \advance\dimen@-#3\dp\tw@ \dimen@ii#2\wd\z@
 \raise-\dimen@\hbox to\dimen4{%
 \hss\kern\dimen@ii\box\tw@\kern-\dimen@ii\hss}%
 \llap{\hbox to\dimen4{\hss\box\z@\hss}}}}

\def\slashed#1{%
 \expandafter\ifx\csname sla@\string#1\endcsname\relax
{\mathpalette{\sla@/00}{#1}}
\fi}
\def\declareslashed#1#2#3#4#5{%
 \expandafter\def\csname sla@\string#5\endcsname{%
#1{\mathpalette{\sla@{#2}{#3}{#4}}{#5}}}}
 \catcode`\@=12
\declareslashed{}{/}{.08}{0}{D}
 \declareslashed{}{/}{.1}{0}{A}
 \declareslashed{}{/}{0}{-.05}{k}
 \declareslashed{}{/}{.1}{0}{\partial}
 \declareslashed{}{\not}{-.6}{0}{f}

\def\lsim{\mathrel {\vcenter {\baselineskip 0pt \kern 0pt
    \hbox{$<$} \kern 0pt \hbox{$\sim$} }}}
\def\gsim{\mathrel {\vcenter {\baselineskip 0pt \kern 0pt
    \hbox{$>$} \kern 0pt \hbox{$\sim$} }}}

\newcommand{\bea}{\begin{eqnarray}}
\newcommand{\eea}{\end{eqnarray}}

\begin{document}

\baselineskip=15pt
\preprint{}

\title{Top-quark forward-backward asymmetry from a color-octet $t$-channel resonance}

\author{Li Cheng$^{1}$\footnote{Electronic address: lcheng@iastate.edu}, Alper Hayreter$^{2}$\footnote{Electronic address: alper.hayreter@ozyegin.edu.tr} and German Valencia$^{1}$\footnote{Electronic address: valencia@iastate.edu}}

\affiliation{$^{1}$ Department of Physics, Iowa State University, Ames, IA 50011.}

\affiliation{$^{2}$ Department of Natural and Mathematical Sciences, Ozyegin University, 34794 Istanbul Turkey.}

\date{\today}

\vskip 1cm
\begin{abstract}

We consider new physics contributions to the top-quark forward-backward asymmetry from a neutral $V^0_8$ or charged $V^+_8$ color-octet vector exchanged in the $t$-channel. We study the phenomenological constraints on these particles arising from the Tevatron and LHC7 measurements and compare them with those on their color singlet counterparts $Z^\prime$ and $W^\prime$. We find that the color octets fare better than the singlets in that they generate a lower $A_C$,  a lower high-invariant mass cross-section at LHC7 and  a lower same sign top-pair cross-section. However, they also generate a lower $A_{FB}$ than their color-singlet counterparts. 

\end{abstract}

\pacs{PACS numbers: }

\maketitle

\section{Introduction}

The forward-backward asymmetry was first observed in top-quark pair production at the Tevatron D0 experiment, $A_{FB}=(19.6\pm 6.5)\%$ \cite{Abazov:2007ab}, to be larger than the standard model (SM) expectation of around 6\% \cite{Kuhn:1998jr,Kuhn:1998kw}. Although this effect is only at the two standard deviation level, it remains a hint for possible new physics after improved measurements and calculations.

Since the first D0 result, the asymmetry measurement has been repeated by both D0 and CDF with increased luminosity with the latest CDF result being obtained with 9.4 fb$^{-1}$ \cite{Aaltonen:2012it}. The corresponding theoretical predictions have been improved to beyond NLO \cite{Ahrens:2011uf} with the observation remaining about two sigma above the SM prediction. This situation has produced a large number of papers exploring the possibility of a new physics explanation for the deviation. Among the first possibilities considered was an axigluon \cite{Ferrario:2008wm,Ferrario:2009bz,Frampton:2009rk,Barcelo:2011fw,Haisch:2011up,Barcelo:2011vk,Tavares:2011zg,Alvarez:2011hi,Drobnak:2012cz} for which a window in the light mass region remains a viable option \cite{Gresham:2012kv}. Many other models have been discussed in this context, including extra dimensions \cite{Delaunay:2011vv,Djouadi:2009nb,Djouadi:2011aj}; composite models \cite{Burdman:2010gr,Alvarez:2010js}; models with $Z^\prime$ bosons \cite{Jung:2009jz,Jung:2011zv,Cao:2011ew,Bhattacherjee:2011nr,Gupta:2010wx,AguilarSaavedra:2011zy,Berger:2011ua,Barreto:2011au,Ko:2012ud,Drobnak:2012rb,Alvarez:2012ca}; models with $W^\prime$ (or both) bosons \cite{Cheung:2009ch,Cheung:2011qa,Cao:2010zb,Barger:2010mw, Barger:2011ih,Frank:2011rb,Shelton:2011hq}; models with extra scalars \cite{Shu:2009xf,Arhrib:2009hu,Dorsner:2009mq,Dorsner:2010cu,Patel:2011eh,Ligeti:2011vt,Grinstein:2011yv,Grinstein:2011dz,Blum:2011fa}. Model independent analyses in terms of effective operators also exist \cite{Jung:2009pi,Degrande:2010kt,Blum:2011up, Ng:2011jv,Gabrielli:2011jf,Gabrielli:2011zw,Gabrielli:2012pk,Shao:2011wa,Biswal:2012mr}; as well as studies that compare different models and study the implications for observables at LHC \cite{Cao:2009uz,Choudhury:2010cd,Gresham:2011pa,Bai:2011ed,Hewett:2011wz,AguilarSaavedra:2011ug,AguilarSaavedra:2011hz,AguilarSaavedra:2011vw,Delaunay:2011gv,Westhoff:2011tq,AguilarSaavedra:2012ma,AguilarSaavedra:2012va}.

A recent comparison of models has found that it is very hard for simple models (those consisting of the exchange of one new particle) to satisfy all existing constraints from cross-sections and asymmetries at the Tevatron and at the LHC \cite{Aguilar-Saavedra:2013rza}. Amongst these simple models there is one case that has not been studied in detail before, new vector color octet particles exchanged in the $t$-channel. Our purpose in this paper is to consider this case, comparing our results to the color-singlet counterparts $Z^\prime$ and $W^\prime$. More complicated models considered before in Ref.~\cite{Grinstein:2011dz}, include color-octet vectors that can be exchanged in the $t$-channel. But this particular effect exists in isolation within that model only for the charged vector case. Ref.~\cite{Gresham:2011fx} also presented a catalog of possible resonances contributing to $A_{FB}$, of which a neutral color-octet vector in the $t$-channel is a possibility. Neither one of these papers presents a comprehensive study of the effect of color-octet resonances in the $t$-channel, which we do in this paper at the {\tt MadGraph5}  level.

\section{Observables}

The original discrepancy with the SM prediction was observed at the Tevatron in the top-quark forward-backward asymmetry
\begin{eqnarray}
A_{FB}&=& \frac{N(\Delta y > 0)-N(\Delta y < 0)}{N(\Delta y > 0)+N(\Delta y < 0)}
\end{eqnarray}
where $\Delta y = y_t- y_{\bar{t}}$ is the difference between the rapidities of the top quark and antiquark with the $z$-axis taken along the proton direction. This asymmetry is equivalent to the top-quark forward-backward asymmetry in the top-quark production angle in the  $t\bar{t}$ rest frame. The asymmetry has now been measured repeatedly  by both D0 and CDF  \cite{Abazov:2007ab,Aaltonen:2008hc,Aaltonen:2011kc,Abazov:2011rq,Aaltonen:2012it} with results that have been consistently above the SM expectation. Within the SM, the asymmetry originates through QCD interference effects at order ${\cal O}(\alpha_s^3)$. It was first predicted by Kuhn and Rodrigo \cite{Kuhn:1998jr,Kuhn:1998kw,Hollik:2011ps,Bernreuther:2012sx,Skands:2012mm,Hoeche:2013mua} and has been revisited several times since then \cite{Bowen:2005ap,Almeida:2008ug,Ahrens:2011mw,Kuhn:2011ri}, including beyond NLO analysis \cite{Ahrens:2011uf}.
For our study we will use the latest CDF results available \cite{Aaltonen:2012it}, as well as the theory prediction quoted by CDF as obtained using the NLO event generator POWHEG,
\begin{eqnarray}
A_{FB}&=&(16.4\pm 4.7)\% \nonumber \\
A_{FB}&=&(6.6\pm 2.0)\%{\rm ~POWHEG}
\label{cdfres}
\end{eqnarray}

We will assume that potential new physics contributions are small, as supported by the agreement between theory and experiment for the $t\bar{t}$ production cross-section, for example. In this case any new physics contributions to the asymmetry can be treated at leading order and simply added to the SM result. The numbers in Eq.~\ref{cdfres} then allow for a new physics contribution to the asymmetry, adding all errors in quadrature,
\begin{eqnarray}
0.05 < &A_{FB}^{new}& < 0.15.
\label{afbnew}
\end{eqnarray}

In addition to the integrated (over $t\bar{t}$ invariant mass) forward backward asymmetry, the Tevatron experiments have also measured an approximately linear dependence of $A_{FB}$ on $m_{tt}$. As a second observable to constrain new physics scenarios we adopt the high invariant mass asymmetry as reported by CDF and the corresponding theoretical prediction quoted by them  in Ref.~\cite{Aaltonen:2012it}, 
\begin{eqnarray}
A_{FB}(M_{t\bar{t}}\geq450~{\rm GeV})&=&(29.5\pm 5.8\pm3.3)\% \nonumber \\
A_{FB}(M_{t\bar{t}}\geq450~{\rm GeV})&=&(10.0\pm 3.0)\% {\rm ~POWHEG}.
\label{cdfhigh}
\end{eqnarray}
Again this leaves room for a new physics contribution after adding all errors in quadrature,
\begin{eqnarray}
0.12 < &A_{FB}^{new}(M_{t\bar{t}}\geq450~{\rm GeV})& < 0.27
\label{afbhighnew}
\end{eqnarray}

The total $t\bar{t}$ production cross-section can also provide a powerful constraint on new physics given the good agreement between the measured and SM predicted values. The combined D0 and CDF results are given in Ref.~\cite{Aaltonen:2013wca} which also quotes the corresponding SM result at NNLO+NNLL QCD based on Ref.~\cite{Czakon:2013goa} for $m_t=172.5$~GeV,
\begin{eqnarray}
\sigma &=& 7.35^{+0.28}_{-0.33}  {\rm ~pb ~~SM} \nonumber \\
\sigma &=& (7.60 \pm 0.41)  {\rm ~pb~~D0-CDF~combination} 
\end{eqnarray}
Adding all errors in quadrature, this allows a new physics contribution to the $p\bar{p}\to t \bar{t}$ cross-section at Tevatron energies
\begin{eqnarray}
\sigma -\sigma_{SM}&=& (0.25 \pm 0.5) {\rm ~pb}.
\label{sigtev1}
\end{eqnarray}

As first noted in Ref.~\cite{Kuhn:1998jr,Kuhn:1998kw}, it is possible to define a related charge asymmetry for proton proton colliders.  Taking advantage of the larger average valence quark momentum than the average  anti-quark momentum in a proton,  one of the observables that have been proposed for the LHC is a charge asymmetry defined by
\begin{eqnarray}
A_{C}&=& \frac{N(\Delta |y| > 0)-N(\Delta |y| < 0)}{N(\Delta |y| > 0)+N(\Delta |y| < 0)},
\end{eqnarray}
where $\Delta |y| = |y_t|- |y_{\bar{t}}|$. This asymmetry has been measured both by CMS and ATLAS with somewhat different results, but in agreement with the SM. The CMS result compared to its SM prediction is \cite{CMS:2012nol}
\begin{eqnarray}
A_{C} &=& 0.004 \pm 0.010 \pm 0.012 \nonumber \\
A_{C} &=& 0.0115 \pm 0.0006 {\rm ~SM~here~POWHEG}
\end{eqnarray}
where the first error is statistical and the second systematic. If we again add all errors in quadrature, this leaves room for a new physics contribution to the charge asymmetry
\begin{eqnarray}
-0.023 <A_C^{new}< 0.008
\label{acnewcms}
\end{eqnarray}
The corresponding result from ATLAS \cite{ATLAS:2012sla} is
\begin{eqnarray}
A_{C} &=& 0.057 \pm 0.024 \pm 0.015 \nonumber \\
A_C &=& 0.006 \pm 0.002  {\rm ~SM~here}
\end{eqnarray}
which allows a new physics contribution
\begin{eqnarray}
0.023 <A_C^{new}< 0.08
\label{acnewatlas}
\end{eqnarray}
Other related asymmetries have been proposed and measured at the LHC but we will only use $A_C$ for the reconstructed $t\bar{t}$ pair in this paper. As with the Tevatron, the total cross-section also places severe constraints on potential new physics. The theoretical \cite{Czakon:2013goa}, ATLAS \cite{ATLAS:2012fja} and CMS \cite{Chatrchyan:2012bra} numbers for 7 TeV collisions at the LHC are respectively given by,
\begin{eqnarray}
\sigma &=& 172  {\rm ~pb~~SM} \nonumber \\
\sigma &=&(177 \pm 3^{+8}_{-7}  \pm 7 )  {\rm ~pb~~ATLAS} \nonumber \\
\sigma &=&(161.9\pm 2.5^{+5.1}_{-5.0} \pm 3.6)  {\rm ~pb~~CMS}
\end{eqnarray}
The theory uncertainties from scale dependence and parton distribution functions estimated at about 3\% each. The CMS  and ATLAS uncertainties quoted correspond to statistical, systematic and luminosity in that order. 
Adding all errors in quadrature and using the ATLAS result  allows a new physics contribution to the cross-section\begin{eqnarray}
\sigma -\sigma_{SM} \leq 18 {\rm ~pb}.
\label{siglhc1}
\end{eqnarray}
The CMS result, being below the SM prediction, would constrain new physics contributions to the cross-section much more severely, our results using Eq.~\ref{siglhc1} would be in agreement with the CMS result at about $\sim 3 \sigma$ level. 

In addition to the total cross-section, it has been pointed out that the high tail cross-section at the LHC provides another constraint on new physics \cite{AguilarSaavedra:2011vw,Delaunay:2011gv}. In Ref.~\cite{Aad:2012hg} ATLAS reported that in the combined lepton plus jets channels at 7 TeV with 2.05 fb$^{-1}$ the high mass tail cross-section 
\begin{eqnarray}
\frac{\sigma(pp\to t\bar{t})(m_{t\bar{t}} > 950 {\rm ~GeV})}{\sigma(pp\to t\bar{t})}= (1.2\pm0.5)\%,
\label{shlhc}
\end{eqnarray}
about 1~$\sigma$ below the theoretical prediction. This has been converted into a 99\% c.l upper limit $\sigma/\sigma_{SM} \leq 1.3$ in this mass bin in Ref. \cite{Aguilar-Saavedra:2013rza}. The CMS result is a bit higher than the ATLAS result but uses different binning \cite{Chatrchyan:2012saa}. 

Finally, for the case of a neutral new particle such as a $Z^\prime$, it is known that there is a strong constraint from same charge top-pair production at the LHC \cite{Gupta:2010wx,AguilarSaavedra:2011zy,Berger:2011ua}. The experimental limit from LHC7 at 95\% c.l. for the inclusive cross-section from ATLAS is  
\begin{equation}
\sigma(pp \to tt) < 4 {\rm ~pb}
\label{sigttlim}
\end{equation}
and the limit from CMS is weaker \cite{Yazgan:2013pxa}.

\section{Vector color octets in the $t$-channel}

As already mentioned, our aim in this paper is to complete the picture of simple new physics contributions to the top-quark forward backward asymmetry by considering the $t$ channel exchange of spin one color octet resonances. These resonances may arise, for example, in techicolor models as color-octet neutral or charged technirhos \cite{Farhi:1980xs,Eichten:1984eu,Lane:2002sm}. The origin of these resonances, however, is not relevant for our phenomenological analysis which will simply follow from the effective Lagrangian
\begin{eqnarray}
{\cal L} &=& -\frac{g_{W_R}}{2} \ \bar{t}\gamma_\mu T_a (1+\gamma_5) \ d \ V^{+\mu}_8 \ - \ \frac{g_{Z_R}}{2} \ \bar{t}\gamma_\mu T_a (1+\gamma_5) \ u \ V^{0\mu}_8 \ + \ {\rm ~h.~c.}
\label{newL8}
\end{eqnarray}
The form chosen is of course motivated by the existing studies of $Z^\prime$ and $W^\prime$ contributions. In particular the flavor structure is chosen to maximize the contribution to the $A_{FB}$, as is the use of right-handed couplings. We will find it useful to compare our results to those of  $Z^\prime$ and $W^\prime$ models in the form
\begin{eqnarray}
{\cal L} &=& -\frac{g_{W_R}}{2} \ \bar{t}\gamma_\mu  (1+\gamma_5) \ d \ V^{+\mu} \ - \ \frac{g_{Z_R}}{2} \ \bar{t}\gamma_\mu  (1+\gamma_5) \ u \ V^{0\mu} \ + \ {\rm ~h.~c.}
\label{newL0}
\end{eqnarray}
It is interesting to note that a recent study of new physics contributions to $A_{FB}$ in terms of effective four-fermion operators arising from $s$-channel exchanges of new particles finds that color octet structures provide a better fit to the data \cite{Gripaios:2013rda}. That analysis does not cover our model of Eq.~\ref{newL8} because the color structure arising from the $t$-channel exchange of the new vectors does not appear in their basis of effective operators. Our analysis of the new physics contributions will only be at leading order, thus ignoring interference between the new physics and NLO SM where the color structure could play an important role. It nevertheless differs from the color singlet structure of $Z^\prime$ and $W^\prime$ models as in Eq.~\ref{newL0} because the different color structure gives a different sign to the interference terms in the differential cross-section. Explicitly, the terms corresponding to the parton level process $q\bar{q} \to t\bar{t}$ including the dominant SM gluon exchange amplitude and the exchange of a $t$-channel $V^{0,+}$color octet or singlet are given by 
\begin{eqnarray}
\frac{d\sigma}{dt} &=& C_{f1}\ \frac{g_s^4}{8\pi s^4}\left(2sm_t^2+(t-m_t^2)^2+(u-m_t^2)^2\right) \nonumber \\
&+&  C_{f3}\ \frac{g_V^4}{16\pi s^2(t-m_{V}^2)^2}\left((u-m_t^2)^2(L_{V}^4+R_V^4)-2L_V^2R_V^2s(t+u)\right)
\nonumber\\
&-& C_{f2}\ \frac{g_V^2g_s^2}{8\pi s^3(t-m_{V}^2)}\left((u-m_t^2)^2+m_t^2s)(L_V^2+R_V^2)\right)
\label{anres}
\end{eqnarray}
where $g_V$ and $m_V$ refer to either $g_{W_R}$ and $m_{W_R}$ or to $g_{Z_R}$ and $m_{Z_R}$. For purely right-handed couplings as in Eq.~\ref{newL8} or in  Eq.~\ref{newL0} $R_V=1$ and $L_V=0$. The color factors for the case of color singlet resonances are given by
\begin{eqnarray}
2 C_{f1}=C_{f2}=\frac{4}{9},&& C_{f3}=1;
\end{eqnarray}
and for color octet resonances by
\begin{eqnarray}
C_{f1}=C_{f3}=\frac{2}{9},&& C_{f2}=-\frac{2}{27}.
\end{eqnarray}
The different color factors are responsible for the different weights of the new physics contributions, including a possible sign change. 

\section{Numerical Results}

For our numerical study we implement the Lagrangian of Eq.~\ref{newL8} and Eq.~\ref{newL0} into {\tt MadGraph5} \cite{MadGraph} with the aid of {\tt FeynRules} \cite{Christensen:2008py}. We use the resulting model UFO file to generate top-quark pair events for different values of couplings and masses in a range that roughly reproduces the new contribution to $A_{FB}$ as in Eq.~\ref{afbnew}. We study the different observables discussed above for these ranges and compare the cases of color octet and color singlet resonances. Finally we fit the numerical results for the case of $m_V = 500$~GeV to obtain approximate expressions for cross-sections and asymmetries in terms of  the new couplings. The results of this fit are presented in the Appendix. 

In Figure~\ref{sigtev} we plot the deviation in the Tevatron $t\bar{t}$ cross-section from its SM value as a function of $A_{FB}^{new}$ for one resonance at a time. The range we show is limited by the charged color octet $V^+_8$ which does not produce a very large $A_{FB}$, the points shown corresponding to $1.4 < g_{W_R} <2$. To obtain a similar asymmetry with $V^0_8$ we show the  range $0.8 < g_{Z_R} < 1.2$. The correlation between cross-section and asymmetry exhibited in Figure~\ref{sigtev} for the color-octets shows that $A_{FB}$ can't be much larger than about $7\%$ if the cross-section is to remain within the 1$\sigma$ range of Eq.~\ref{sigtev1}. In the same figure we show in the right panel the case of color singlet resonances using the ranges $0.5 < g_{W_R} <2$ and $1.15 < g_{Z_R} <1.4$. The color singlets allow much larger asymmetries due to the relatively larger color factor in the interference term in  Eq.~\ref{anres}. The corresponding cross-sections are also lower for the singlet as there is destructive interference with the SM. In fact all the points shown in Figure~\ref{sigtev} for $V^0$ (color singlet) have a cross-section below the 1$\sigma$ bound of Eq.~\ref{sigtev1}.
\begin{figure}[htb]
\includegraphics[width=0.46\textwidth]{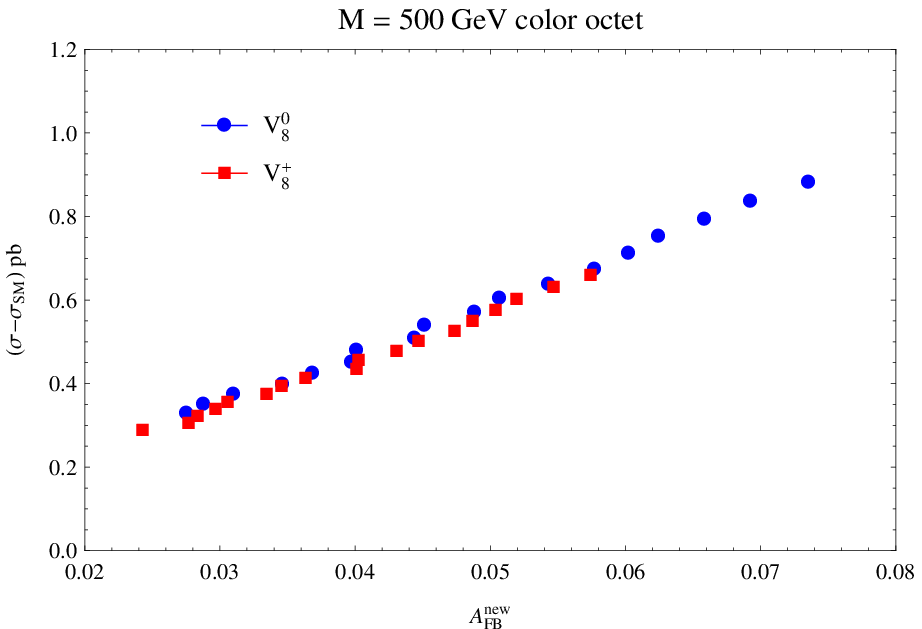}
\includegraphics[width=0.46\textwidth]{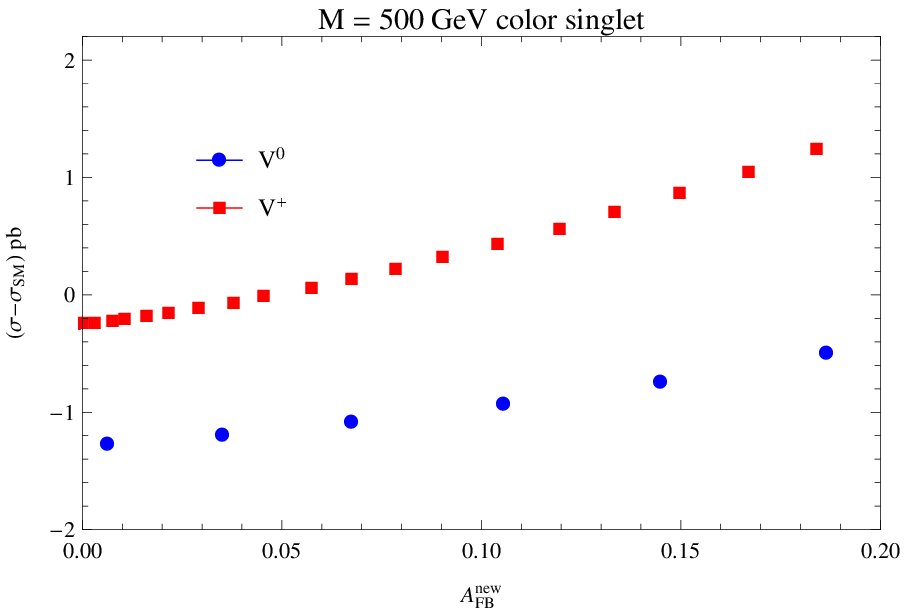}
\caption{Deviation in the Tevatron cross-section from its SM value as a function of $A_{FB}^{new}$ for one resonance at a time.}
\label{sigtev}
\end{figure}

\begin{figure}[htb]
\includegraphics[width=0.46\textwidth]{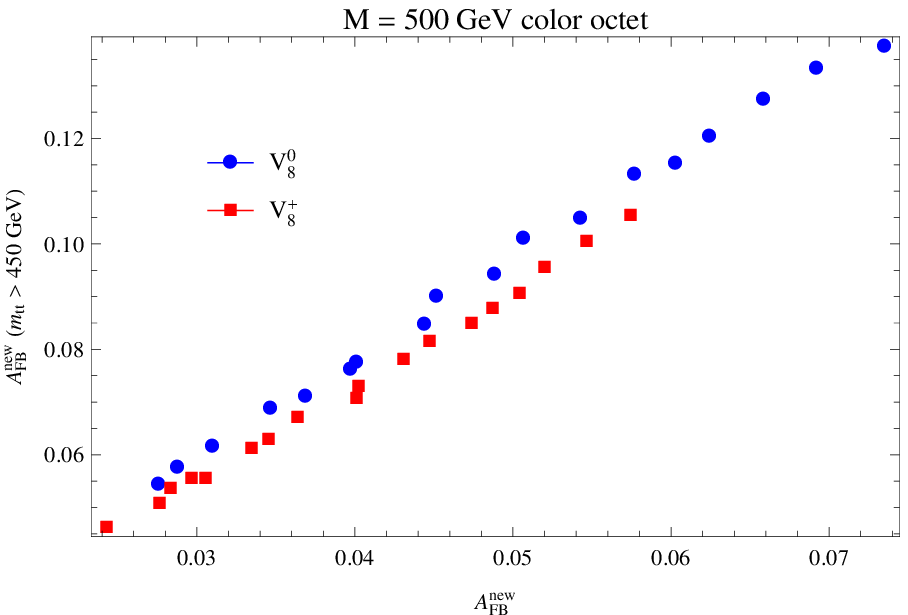}
\includegraphics[width=0.46\textwidth]{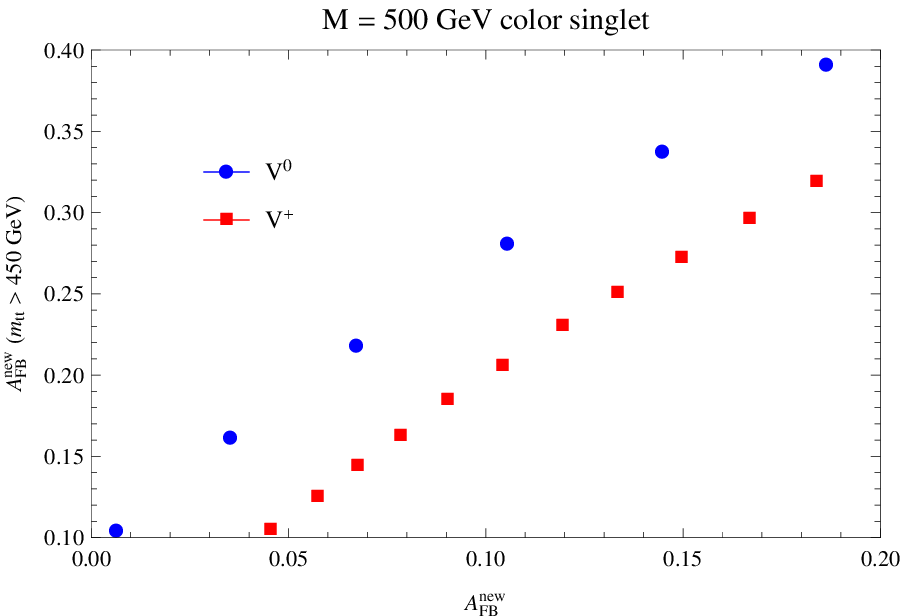}
\caption{$A_{FB}^{new}(m_{t\bar{t}}>450 {\rm ~GeV})$ as a function of $A_{FB}^{new}$ for one resonance at a time.}
\label{ahtev}
\end{figure}

In Figure~\ref{ahtev} we plot the high invariant mass Tevatron asymmetry $A_{FB}^{new}(m_{t\bar{t}}>450 {\rm ~GeV})$ as a function of $A_{FB}^{new}$ for one resonance at a time using the same couplings as in Fig.~\ref{sigtev}. The results indicate that the color-octet resonances can only reproduce the lower ends of the 1$\sigma$ ranges of Eqs.~\ref{afbnew}~and~\ref{afbhighnew}. The right panel, corresponding  to the color singlets, corroborates that a $Z^\prime$  tends to over-predict the high invariant mass asymmetry \cite{AguilarSaavedra:2011hz}.

\begin{figure}[htb]
\includegraphics[width=0.46\textwidth]{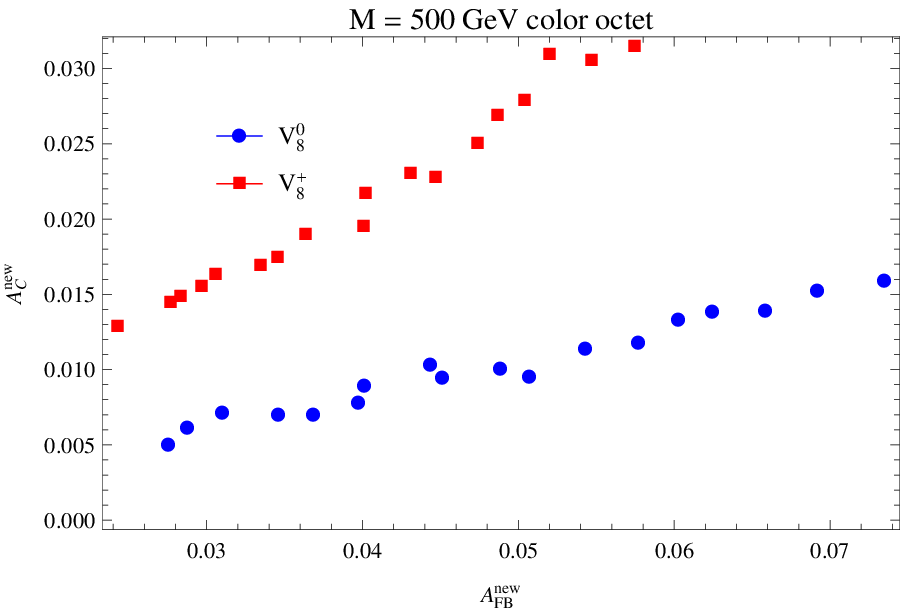}
\includegraphics[width=0.46\textwidth]{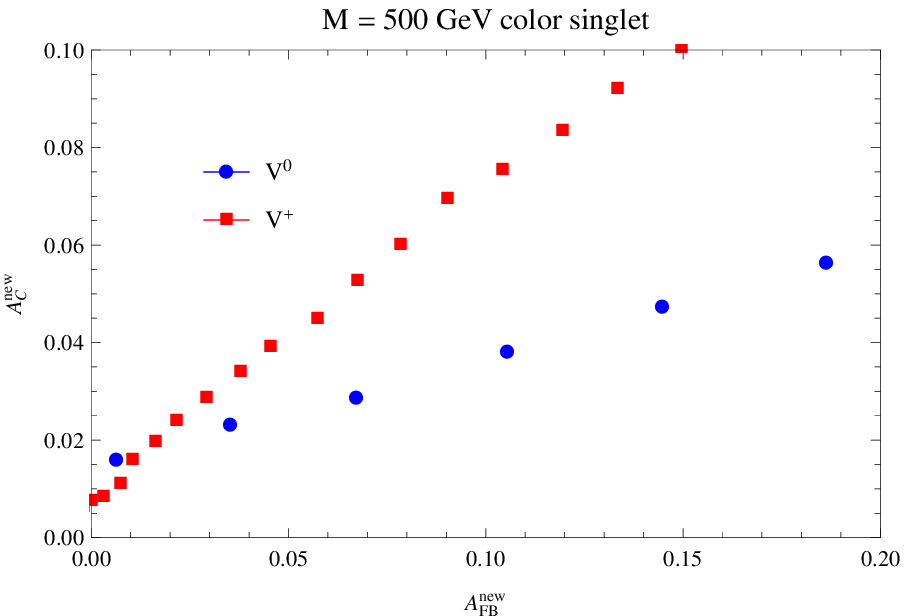}
\caption{$A_{C}^{new}$  as a function of $A_{FB}^{new}$  for one resonance at a time}
\label{aclhc}
\end{figure}

In Figure~\ref{aclhc} we plot the charge asymmetry $A_{C}^{new}$ for LHC7 as a function of $A_{FB}^{new}$ for one resonance at a time with the same couplings used in Fig.~\ref{sigtev}. The correlation between these two observables is such that a neutral boson is preferred over a charged boson by the measured $A_C$. In fact,  the tighter CMS constraint from Eq.~\ref{acnewcms} at the 1$\sigma$ level, only allows the neutral color-octet. The panel on the right, again for the color singlets, corroborates that current LHC data  disfavors a $W'$ as it over-predicts $A_C$ \cite{AguilarSaavedra:2011hz,Aguilar-Saavedra:2013rza}.

\begin{figure}[htb]
\includegraphics[width=0.46\textwidth]{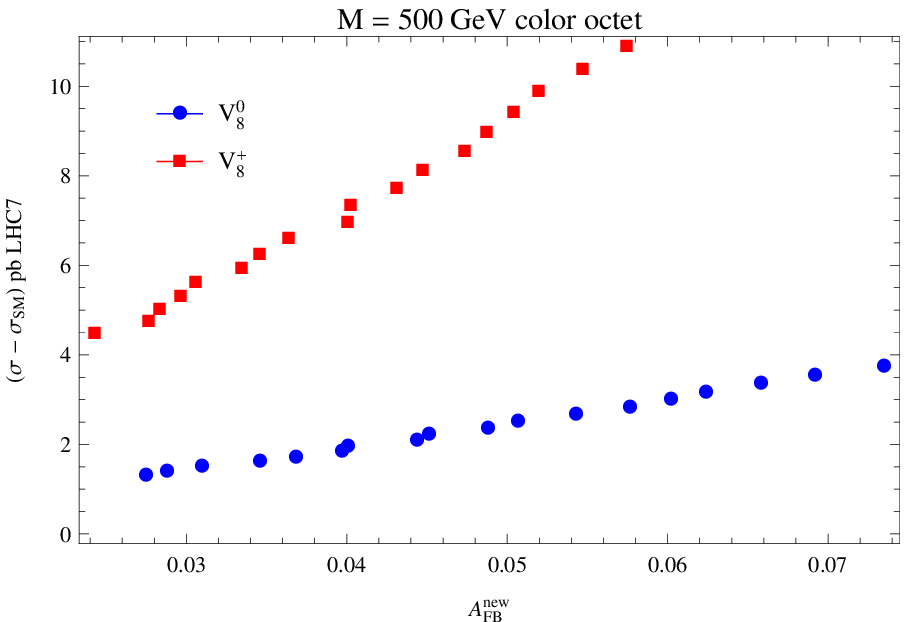}
\includegraphics[width=0.46\textwidth]{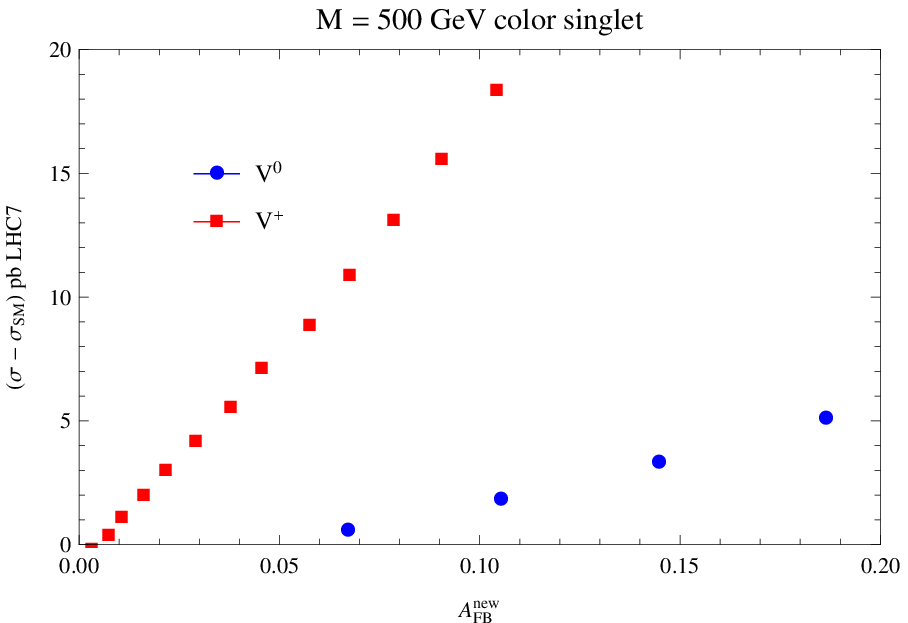}
\caption{Deviation in the LHC7  cross-section from its SM value as a function of $A_{FB}^{new}$ for one resonance at a time.}
\label{siglhc}
\end{figure}

In Figure~\ref{siglhc} we plot the deviation in the LHC7  cross-section from its SM value as a function of $A_{FB}^{new}$ for one resonance at a time with the same couplings used in Fig.~\ref{sigtev}. All the points shown satisfy the 1$\sigma$ range from Eq.~\ref{siglhc1} obtained from the ATLAS result but the $W^\prime$ and its color octet counterpart $V^+_8$ give the largest cross-sections, possibly in conflict with the CMS measurement.

\begin{figure}[thb]
\includegraphics[width=0.46\textwidth]{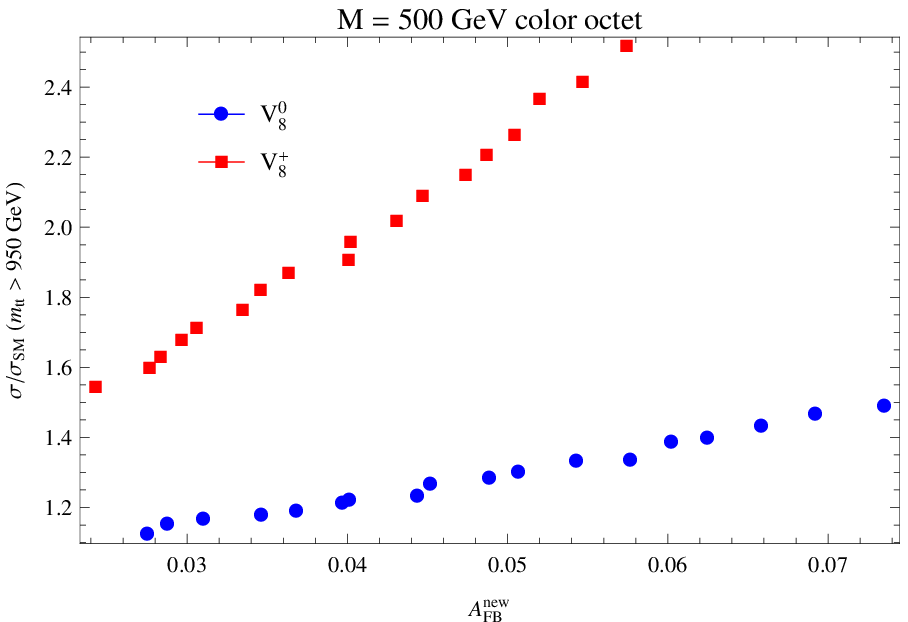}
\includegraphics[width=0.46\textwidth]{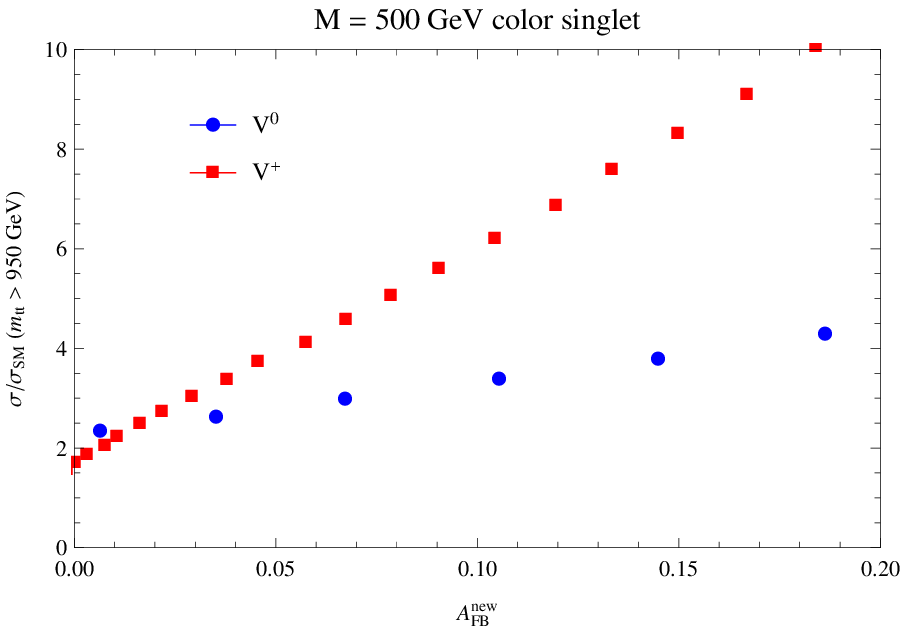}
\caption{High invariant mass cross-section at LHC7 as a function of $A_{FB}^{new}$ for one resonance at a time}
\label{sighlhc}
\end{figure}

In Figure~\ref{sighlhc} we plot the high invariant mass cross-section at LHC7 as a function of $A_{FB}^{new}$ for one resonance at a time using the same couplings as in Fig.~\ref{sigtev}. Again the neutral resonances fare better than the charged ones and the color-octet much better than the color singlet in a comparison with Eq.~\ref{shlhc}.

\begin{figure}[htb]
\includegraphics[width=0.48\textwidth]{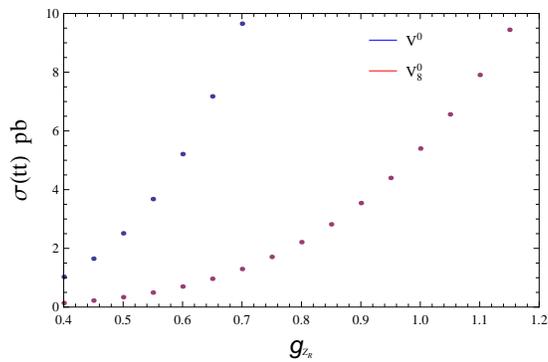}
\caption{$\sigma(pp\to tt)$ at LHC7 for the neutral bosons of mass 500~GeV.}
\label{ttpairs}
\end{figure}

Finally in Figure~\ref{ttpairs} we show the cross-section for double top-quark production, $\sigma(pp\to tt)$, at LHC7 for the neutral bosons of mass 500~GeV. The figure indicates that the color singlet $Z^\prime$ quickly runs into trouble with the ATLAS limit on this process, Eq.~\ref{sigttlim}, but the color octet fares better.

\begin{figure}[htb]
\includegraphics[width=0.48\textwidth]{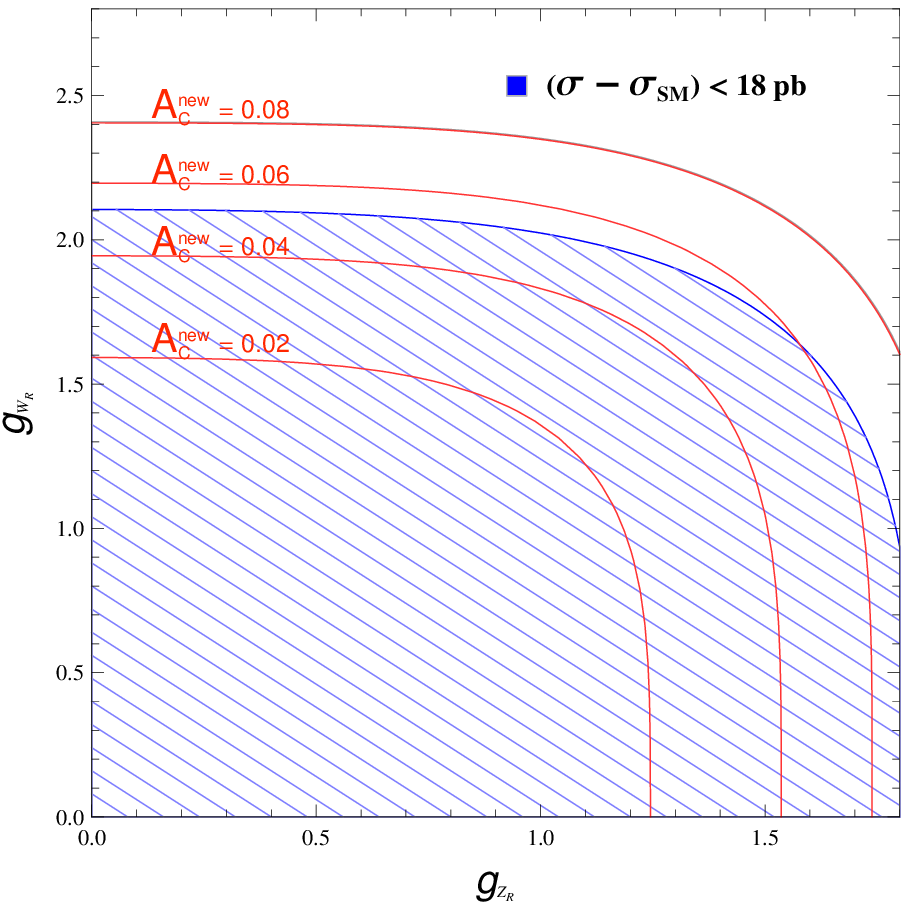}
\includegraphics[width=0.48\textwidth]{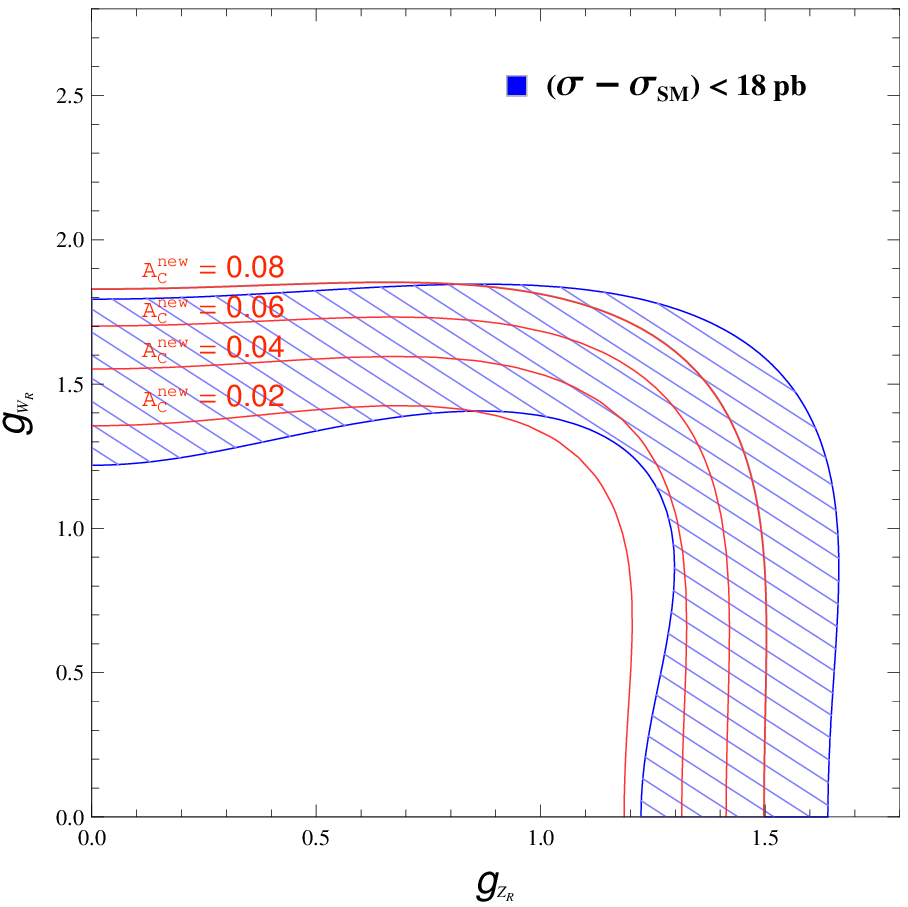}
\caption{Color-octet (left panel) vs color-singlet (right panel) parameter space for couplings allowed by the LHC7 cross-section and charge asymmetry.}
\label{pspace3}
\end{figure}

We now turn to the question of whether there is an optimal region in parameter space to satisfy all the constraints. To this end we use the approximate fits presented in the Appendix to produce 
Figures~\ref{pspace1},~\ref{pspace2} and ~\ref{pspace3} where we compare the allowed regions in the $g_{W_R}-g_{Z_R}$ plane for the different observables discussed above. 
\begin{figure}[htb]
\includegraphics[width=0.48\textwidth]{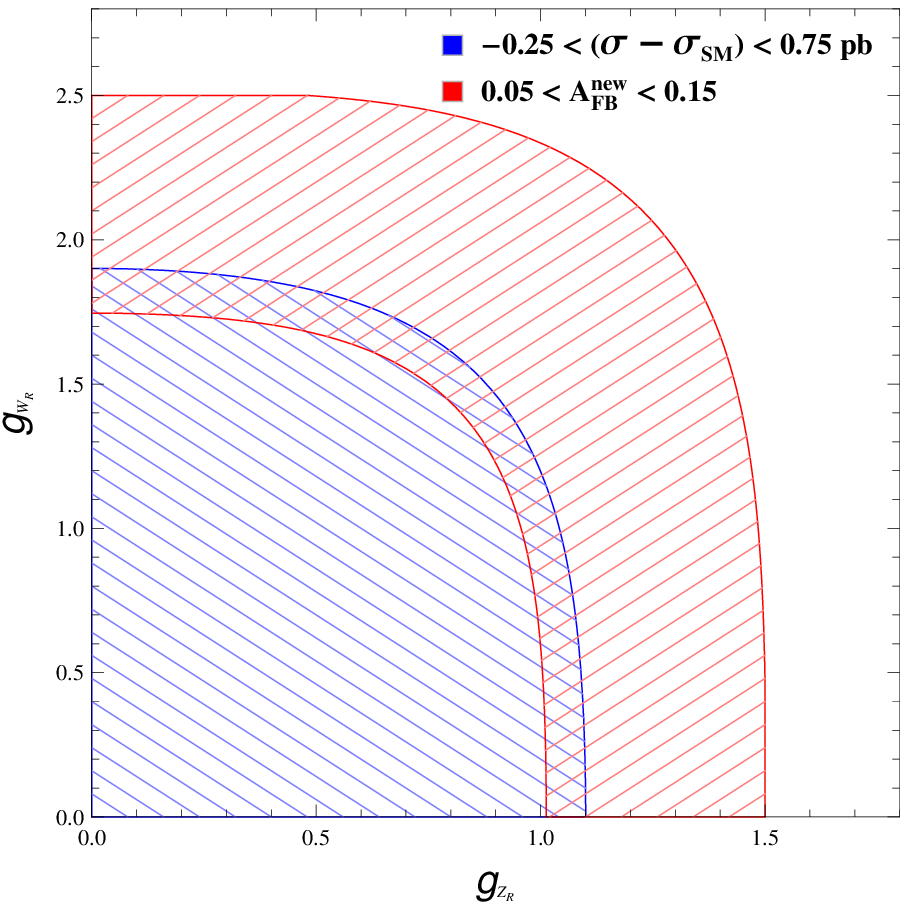}
\includegraphics[width=0.48\textwidth]{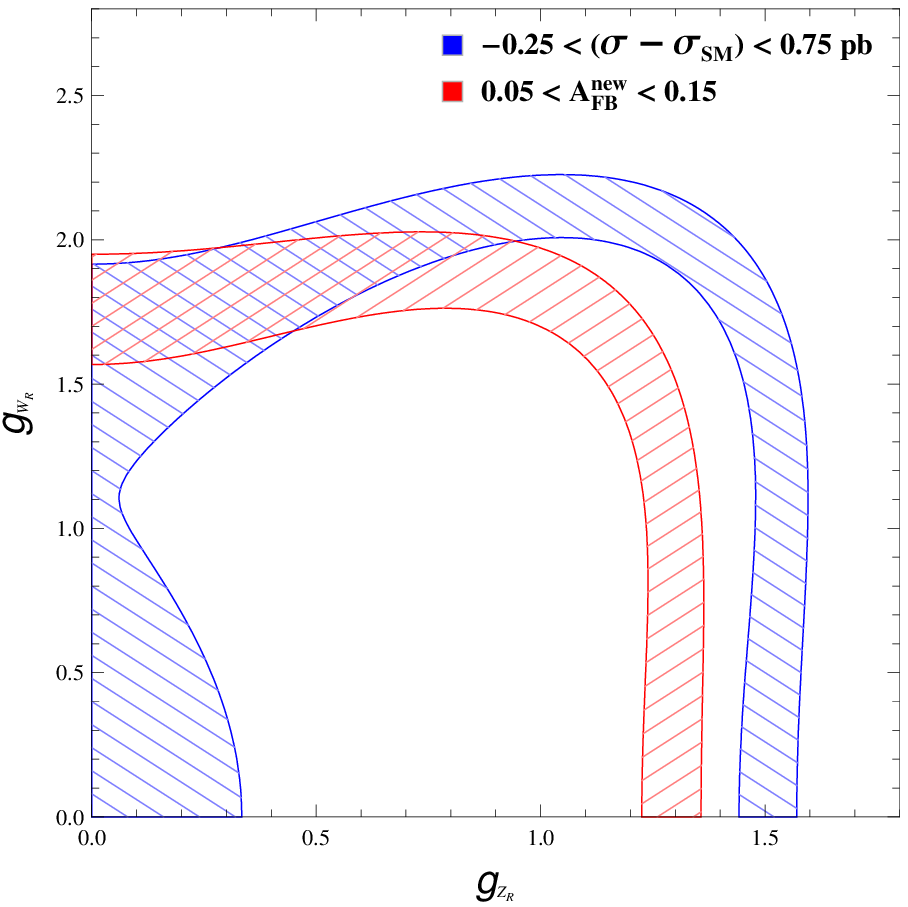}
\caption{Color-octet (left panel) vs color-singlet (right panel) parameter space for couplings allowed by the Tevatron cross-section and forward-backward asymmetry.}
\label{pspace1}
\end{figure}
Figure~\ref{pspace1} shows a cross-section and an asymmetry  at the Tevatron that are compatible at the one-sigma level with a narrow band of parameter space for the color-octet. The new physics required to increase the asymmetry also increases the cross-section and the two are compatible only for the lower end of the 1$\sigma$ range for $A_{FB}^{new}$. This situation is different from the color-singlet where there is destructive interference between the SM and the new physics.

\begin{figure}[htb]
\includegraphics[width=0.48\textwidth]{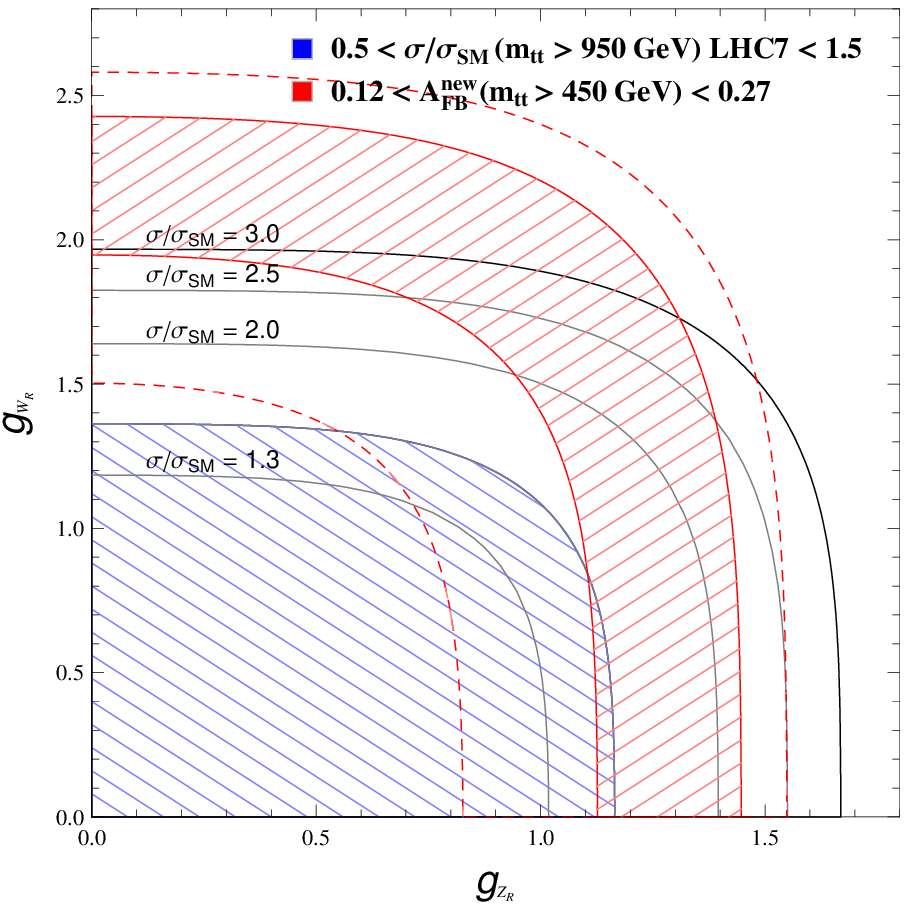}
\includegraphics[width=0.48\textwidth]{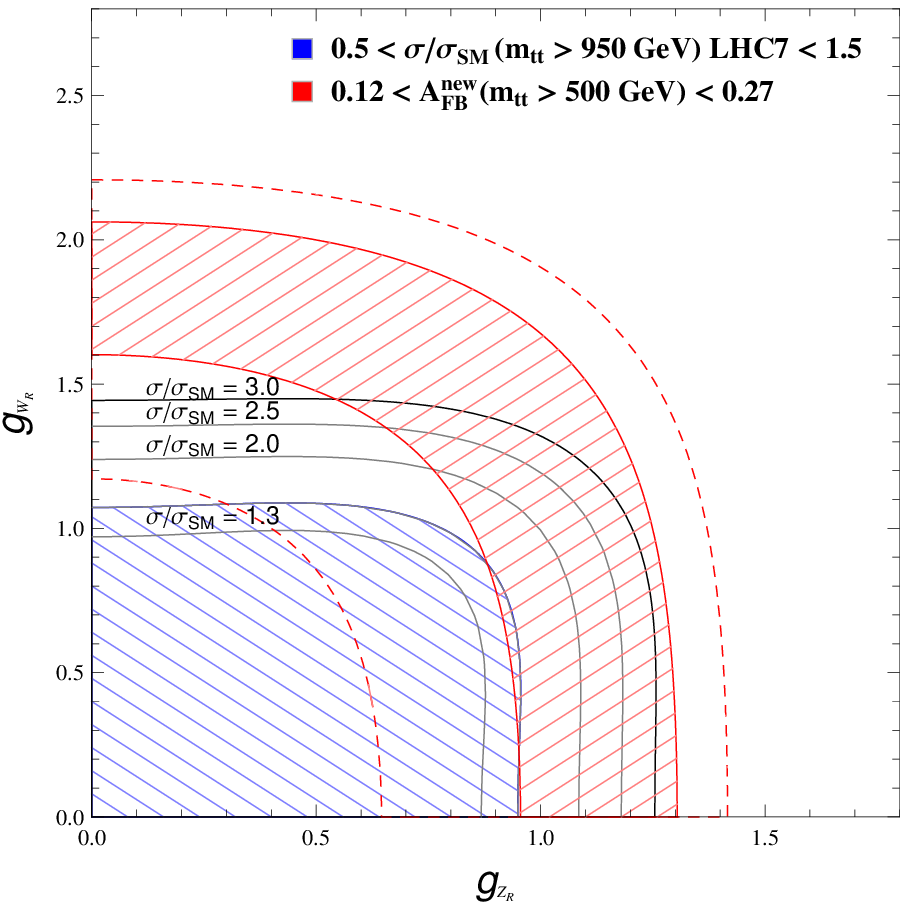}
\caption{Color-octet (left panel) vs color-singlet (right panel) parameter space for couplings allowed by the high invariant mass LHC7 cross-section and high invariant mass Tevatron forward-backward asymmetry. The black lines are contours for higher values of $\sigma(m_{t\bar{t}}>950 {\rm ~GeV})$ at LHC7 than allowed by the current ATLAS measurement. The dashed red lines illustrate the 2$\sigma$ contour for $A_{FB}^{new}(m_{t\bar{t}}>450 {\rm ~GeV})$.}
\label{pspace2}
\end{figure}
In Figure~\ref{pspace2} we examine the effect of the high invariant mass observables. The 99\% c.l upper limit $\sigma/\sigma_{SM}(m_{t\bar{t}}>950 {\rm ~GeV}) \leq 1.3$ from ATLAS quoted in Ref. \cite{Aguilar-Saavedra:2013rza} ruling out both the color-singlet and color octet-resonances as explanations for $A_{FB}$. We also show in the figure the boundaries corresponding to 
$\sigma/\sigma_{SM}(m_{t\bar{t}}>950 {\rm ~GeV}) = 2.0,\ 2.5, {\rm ~and~} 3.0$ to indicate what would be necessary to be compatible with the current 1$\sigma$ range for $A_{FB}^{new}(m_{t\bar{t}}>450 {\rm ~GeV})$. Of course, a lower value of this high invariant-mass asymmetry (at the two sigma level for example) also opens up the allowed parameter space as indicated by the dashed red lines in the Figure.

In Figure~\ref{pspace3} we consider the constraints from the charge asymmetry and cross-section at LHC7. We have indicated several contours for $A_C^{new}$ to compare with the different results found by ATLAS and CMS.
 
For our numerical study we have used a mass of 500~GeV for the new color-octet boson as an illustration. But we also generated similar samples for 600~GeV and smaller samples for masses ranging between 400~GeV and 1~TeV. For all cases we found that the correlations between the different observables are very similar to those exhibited in Figures~\ref{sigtev}-\ref{sighlhc}. The value of $A_{FB}$ for masses higher than 500 GeV that is obtained keeping the couplings fixed gets smaller with increasing mass. For a mass of 600~GeV, for example, the range of $A_{FB}$ shown in Figures~\ref{sigtev}-\ref{sighlhc} can be covered by increasing the couplings used by about 0.2 in each case. For heavier bosons this becomes harder to do as couplings would move into non-perturbative regimes. By the time masses reach 1~TeV it is only possible to generate very small values of $A_{FB}$. We also simulated events for the benchmark point ($m_V=300$~GeV) for the model ``Vector field $VII_O$'' of reference \cite{Grinstein:2011dz} (corresponding to our $V_8^+$); as well as for the points in Table IV, model C8V of Ref.~\cite{Gresham:2011fx} (corresponding to our $V_8^0$) and we are in rough agreement in these cases.

\section{Conclusion}

We have studied the effect from a neutral $V^0_8$ or charged $V^+_8$ color-octet vector exchanged in the $t$-channel on the  top-quark forward-backward asymmetry. We find that they can modestly increase the SM value of $A_{FB}$, to within the 1$\sigma$ range from the Tevatron measurement. The color-octets fare better than the color-singlets when confronted with other constraints. In particular they generate a lower $A_C$,  a lower high-invariant mass cross-section at LHC7 and a lower same sign top-pair cross-section. We have studied the correlations between the different observables  for a mass of 500~GeV and the corresponding parameter space that is still allowed. We find that this type of new physics is still consistent with the measurements  at the two sigma level.

\begin{acknowledgments}

This work was supported in part by the DOE under contract number DOE under contract number DE-SC0009974. We are grateful to Chunhui Chen for useful discussions, to Kingman Cheung for discussions of the models with a $W^\prime$, to Alex Kagan for clarifications on Ref.~\cite{Grinstein:2011dz} and to Moira Gresham for clarifications on Ref.~\cite{Gresham:2011fx}.

\end{acknowledgments}

\appendix

\section{Approximate results for $m_V = 500$~GeV}

We generated samples of one million events for at least 40 points in $g_{W_R},g_{Z_R}$ parameter space with resonance masses of 500 GeV and widths of 50 GeV (although the precise value of the width is not important for $t$-channel resonances). Using these points we performed a fit to a quartic polynomial in these couplings (of the form that occurs in an analytic calculation) to obtain approximate expressions for the different observables. These expressions were then used in our exploration of parameter space. The results of these fits for color-octet resonances and for Tevatron observables are 
\begin{eqnarray}
\sigma(p\bar{p} \to t \bar{t}) &\approx & (6.06 + 0.325 g_{Z_R}^2 +0.245 g_{Z_R}^4+0.074 g_{W_R}^2 +0.037 g_{W_R}^4 ){\rm ~pb} \nonumber \\
\sigma \cdot A_{FB} &=& (0.012 +0.132  g_{Z_R}^2+0.176  g_{Z_R}^4 +0.028  g_{W_R}^2+0.025g_{W_R}^4){\rm ~pb} \nonumber \\
A_{FB} (M_{t\bar{t}}\geq 450 {\rm ~GeV}) &=& (38.6  g_{Z_R}^2+42.6  g_{Z_R}^4+5.02g_{W_R}^2 +6.86g_{W_R}^4)\times 10^{-3}\nonumber 
\end{eqnarray}
Note that for the case of $A_{FB}$ (but not for $A_{FB}$ at high invariant mass) our fit is for  $A_{FB}$ times the cross-section. The constant term in this expression is the electroweak contribution in the SM as calculated by  {\tt MadGraph 5}.
For for color-octet resonances and LHC observables at a 7~TeV energy we obtain
\begin{eqnarray}
\sigma(p p \to t \bar{t}) &\approx & (96.33 + 1.115 g_{Z_R}^2 +1.245 g_{Z_R}^4+0.877 g_{W_R}^2 +0.719 g_{W_R}^4 ){\rm ~pb} \nonumber \\
\sigma \cdot A_{C} &=& (0.1 +0.261  g_{Z_R}^2+0.632  g_{Z_R}^4 +0.040  g_{W_R}^2+0.290g_{W_R}^4){\rm ~pb} \nonumber \\
\sigma(p p \to t \bar{t}) (M_{t\bar{t}}\geq 950 {\rm ~GeV}) &=& (1.2 + 0.042  g_{Z_R}^2+0.294  g_{Z_R}^4+0.050g_{W_R}^2 +0.147g_{W_R}^4){\rm ~pb} \nonumber
\end{eqnarray}
The results of our fits for color-singlet resonances and for Tevatron observables are 
\begin{eqnarray}
\sigma(p\bar{p} \to t \bar{t}) &\approx & (6.06 -2.423 g_{Z_R}^2 +1.106 g_{Z_R}^4-0.407 g_{W_R}^2 +0.167 g_{W_R}^4 ){\rm ~pb} \nonumber \\
\sigma \cdot A_{FB} &=& (0.012 -1.046  g_{Z_R}^2+0.800  g_{Z_R}^4 -0.158  g_{W_R}^2+0.113g_{W_R}^4){\rm ~pb} \nonumber \\
A_{FB} (M_{t\bar{t}}\geq 450 {\rm ~GeV}) &=& (96.1 g_{Z_R}^2+36.0  g_{Z_R}^4+19.8g_{W_R}^2 +10.1g_{W_R}^4)\times 10^{-3}\nonumber
\end{eqnarray}
For for color-singlet resonances and LHC observables at a 7~TeV energy we obtain
\begin{eqnarray}
\sigma(p p \to t \bar{t}) &\approx & (96.33 - 8.40 g_{Z_R}^2 +5.61 g_{Z_R}^4-4.79 g_{W_R}^2 +3.22 g_{W_R}^4 ){\rm ~pb} \nonumber \\
\sigma \cdot A_{C} &=& (0.1-2.82  g_{Z_R}^2+2.92 g_{Z_R}^4 -1.10  g_{W_R}^2+1.15g_{W_R}^4){\rm ~pb} \nonumber \\
\sigma(p p \to t \bar{t}) (M_{t\bar{t}}\geq 950 {\rm ~GeV}) &=& (1.2 -0.482  g_{Z_R}^2+1.28 g_{Z_R}^4-0.254g_{W_R}^2 +0.675g_{W_R}^4){\rm ~pb} \nonumber
\end{eqnarray}

\end{document}